\documentclass[pre,twocolumn,twoside,byrevtex,superscriptaddress,floatfix]{revtex4-1}

\usepackage [english]{babel}
\usepackage [autostyle, english = american]{csquotes}
\usepackage{indentfirst}
\usepackage{xcolor}
\usepackage[section]{placeins}
\MakeOuterQuote{"}
 
  \newcommand{\tb}{\color{black}}

\usepackage{currfile}
\usepackage{graphicx,epsfig,verbatim,enumerate}
\usepackage{amssymb,amsmath}
\usepackage{ifthen}

\usepackage{longtable}

\usepackage{mathtools}

\newboolean{twocolswitch}

\newcommand{\sindex}[1]{}
\newcommand{\nindex}[1]{}

\newcommand{\www}[1]{\url{#1}}

\usepackage{lettrine}

\usepackage{color}

\setboolean{twocolswitch}{true}

\begin{document}

\title{\protect
Climate change sentiment on Twitter: An unsolicited public opinion poll
}

\author{
\firstname{Emily M.}
\surname{Cody}
}
\email{emily.cody@uvm.edu}
\affiliation{Department of Mathematics \& Statistics,
  Vermont Complex Systems Center,
  Computational Story Lab,
  \& the Vermont Advanced Computing Core,
  The University of Vermont,
  Burlington, VT 05401.}

\author{
\firstname{Andrew J.}
\surname{Reagan}
}
\email{andy@andyreagan.com}
\affiliation{Department of Mathematics \& Statistics,
  Vermont Complex Systems Center,
  Computational Story Lab,
  \& the Vermont Advanced Computing Core,
  The University of Vermont,
  Burlington, VT 05401.}

\author{
\firstname{Lewis}
\surname{Mitchell}
}
\email{lewis.mitchell@adelaide.edu.au}
  \affiliation{School of Mathematical Sciences, The University of Adelaide, SA 5005, Australia}

\author{
\firstname{Peter Sheridan}
\surname{Dodds}
}
\email{peter.dodds@uvm.edu}
\affiliation{Department of Mathematics \& Statistics,
  Vermont Complex Systems Center,
  Computational Story Lab,
  \& the Vermont Advanced Computing Core,
  The University of Vermont,
  Burlington, VT 05401.}

\author{
\firstname{Christopher M.}
\surname{Danforth}
}
\email{chris.danforth@uvm.edu}
\affiliation{Department of Mathematics \& Statistics,
  Vermont Complex Systems Center,
  Computational Story Lab,
  \& the Vermont Advanced Computing Core,
  The University of Vermont,
  Burlington, VT 05401.}

%

\date{\today}

\begin{abstract}
  \protect
  The consequences of anthropogenic climate change are extensively debated through scientific papers, newspaper articles, and blogs.  Newspaper articles may lack accuracy, while the severity of findings in scientific papers may be too opaque for the public to understand.  Social media, however, is a forum where individuals of diverse backgrounds can share their thoughts and opinions.  As consumption shifts from old media to new, Twitter has become a valuable resource for analyzing current events and headline news.  In this research, we analyze tweets containing the word "climate" collected between September 2008 and July 2014. \tb{Through use of a previously developed sentiment measurement tool called the Hedonometer, we determine how collective sentiment varies in response to climate change news, events, and natural disasters.  We find that natural disasters,  climate bills, and oil-drilling can contribute to a decrease in happiness while climate rallies, a book release, and a green ideas contest can contribute to an increase in happiness}.  Words uncovered by our analysis suggest that responses to climate change news are predominately from climate change activists rather than climate change deniers, indicating that Twitter is a valuable resource for the spread of climate change awareness. 
\end{abstract}


\maketitle

\section*{Introduction}
After decades receiving little attention from non-scientists, the impacts of climate change are now widely discussed through a variety of mediums.  Originating from scientific papers, newspaper articles, and blog posts, a broad spectrum of climate change opinions, subjects, and sentiments exist.  Newspaper articles often dismiss or sensationalize the effects of climate change due to journalistic biases including personalization, dramatization and a need for novelty \cite{boykoff2007climate}.  Scientific papers portray a much more realistic and consensus view of climate change.  These views, however, do not receive widespread media attention due to several factors including journal paywalls, formal scientific language, and technical results that are not easy for the general public to understand \cite{boykoff2007climate}.

According to the IPCC Fifth Assessment report, humans are "very likely" (90-100\% probability) to be responsible for the increased warming of our planet \cite{field2014climate}, and this anthropogenic global warming is responsible for certain weather extremes \cite{fischer2015anthro}.  In April 2013, 63\% of Americans reported that they believe climate change is happening.  This number, however, drops to 49\% when asked if climate change is being caused by humans.  The percentage drops again to 38\% when asked if people around the world are currently being harmed by the consequences of climate change \cite{leiserowitz2013climate}.  These beliefs and risk perceptions can vary by state or by county \cite{howe2015geographic}.  By contrast,  97\% of active, publishing, climate change scientists agree that "human activity is a significant contributing factor in changing mean global temperatures" \cite{doran2009examining,anderegg2010expert}.  The general public learns most of what it knows about science from the mass-media \cite{wilson1995mass}.  Coordination among journalists, policy actors, and scientists will help to improve reporting on climate change, by engaging the general public and creating a more informed decision-making process  \cite{boykoff2011speaks}.    

One popular source of climate information that has not been heavily analyzed is social media.  The Pew Research Center's Project for Excellence in Journalism in January of 2009 determined that topics involving global warming are much more prominent in the new, social media \cite{boykoff2011speaks}.    In the last decade, there has been a shift from the consumption of traditional mass media (newspapers and broadcast television) to the consumption of social media (blog posts, Twitter, etc.).  This shift represents a switch in communications from "one-to-many" to "many-to-many" \cite{boykoff2011speaks}.  Rather than a single journalist or scientist telling the public exactly what to think, social media offers a mechanism for many people of diverse backgrounds to communicate and form their own opinions.  Exposure is a key aspect in transforming a social problem into a public issue \cite{dearing1996agenda}, and social media is a potential avenue where climate change issues can be initially exposed. 

Here we study the social media site Twitter, which allows its users 140 characters to communicate whatever they like within a "tweet".  Such expressions may include what individuals are thinking, doing, feeling, etc.  Twitter has been used to explore a variety of social and linguistic phenomena \cite{cao2012whisper, lin2013voices, lin2014rising}, and used as a data source to create an earthquake reporting system in Japan \cite{sakaki2010earthquake}, detect influenza outbreaks \cite{aramaki2011twitter}, and analyze overall public health \cite{paul2011you}.  An analysis of geo-tagged Twitter activity (tweets including a latitude and longitude) before, during, and after Hurricane Sandy using keywords related to the storm is given in \cite{nowcasting}.  They discover that Twitter activity positively correlates with proximity to the storm and physical damage.  It has also been shown that individuals affected by a natural disaster are more likely to strengthen interactions and form close-knit groups on Twitter immediately following the event \cite{Phan11052015}.  Twitter has also been used to examine human sentiment through analysis of variations in the specific words used by individuals.  In \cite{dodds2011temporal}, Dodds et al. develop the "hedonometer", a tool for measuring expressed happiness -- positive and negative sentiment -- in large-scale text corpora.  Since its development, the hedonometer has been implemented in studies involving the happiness of cities and states \cite{mitchell2013geography}, the happiness of the English language as a whole \cite{kloumann2012positivity}, and the relationship between individuals' happiness and that of those they connect with \cite{bliss2012twitter}.

The majority of the topics trending on Twitter are headlines or persistent news \cite{kwak2010twitter}, making Twitter a valuable source for studying climate change opinions.  For example, in \cite{antracking}, subjective vs objective and positive vs negative tweets mentioning climate change are coded manually and analyzed over a one year time period.  \tb{In \cite{williams2015network}, various climate hashtags are utilized to locate pro/denialist communities on Twitter.}  In the present study, we apply the hedonometer to a collection of tweets containing the word "climate".  We collected roughly 1.5 million such tweets from Twitter's gardenhose API (a random 10\% of all messages) during the roughly 6 year period spanning September 14, 2008 through July 14, 2014.  \tb{This time period represents the extent of our database at the time of writing.}  Each collected tweet contains the word "climate" at least once.  \tb{We include retweets in the collection to ensure an appropriately higher weighting of messages authored by popular accounts (e.g. media, government).}  We apply the hedonometer to the climate tweets during different time periods and compare them to a reference set of roughly 100 billion tweets from which the climate-related tweets were filtered.  We analyze highest and lowest happiness time periods using word shift graphs developed in \cite{dodds2011temporal}, and we discuss specific words contributing to each happiness score.          

\section*{Methods}

The hedonometer is designed to calculate a happiness score for a large collection of text, based on the happiness of the individual words used in the text.  The instrument uses sentiment scores collected by Kloumann et al. and Dodds et al.  \cite{kloumann2012positivity,dodds2011temporal}, where 10,222 of the most frequenly used English words in four disparate corpora were given happiness ratings using Amazon's Mechanical Turk online marketplace.  Fifty participants rated each word, and the average rating becomes the word's score.  Each word was rated on a scale from 1 (least happy) to 9 (most happy) based on how the word made the participant feel.  We omit clearly neutral or ambiguous words (scores between 4 and 6) from the analysis.  In the present study, we use the instrument to measure the average happiness of all tweets containing the word "climate" from \tb{September 14, 2008 to July 14, 2014} on the timescales of day, week, and month.  The word "climate" has a score of 5.8 and was thus not included when calculating average happiness.  \tb{For comparison, we also calculate the average happiness score surrounding 5 climate related keywords.}

We recognize that not every tweet containing the word "climate" is about climate change.  Some of these tweets are about the economic, political, or social climate and some are ads for climate controlled cars.  Through manual coding of a random sample of 1,500 climate tweets, we determined that 93.5\% of tweets containing the word "climate" are about the earth's climate or climate change.  We calculated the happiness score for both the entire sample and the sample with the non-earth related climate tweets removed.  The scores were 5.905 and 5.899 respectively, a difference of 0.1\%.  This difference is small enough to conclude that the non-earth related climate change tweets do not substantially alter the overall happiness score.

Based on the happiness patterns given by the hedonometer analysis, we select specific days for analysis using word shift graphs.  We use word shift graphs to compare the average happiness of two pieces of text, by rank ordering the words that contribute the most to the increase or decrease in happiness.  In this research, the comparison text is all tweets containing the word "climate", and the reference text is a random 10\% of all tweets.  Hereafter, we refer to the full reference collection as the "unfiltered tweets".

Finally, we analyze four events including three natural disasters and one climate rally using happiness time series and word shift graphs.  These events include Hurricane Irene (August 2011), Hurricane Sandy (October 2012), a midwest tornado outbreak (May 2013), and the Forward on Climate Rally (February 2013).     

\section*{Results}

\begin{figure*}[tbp!]
\includegraphics[width=0.9\textwidth]{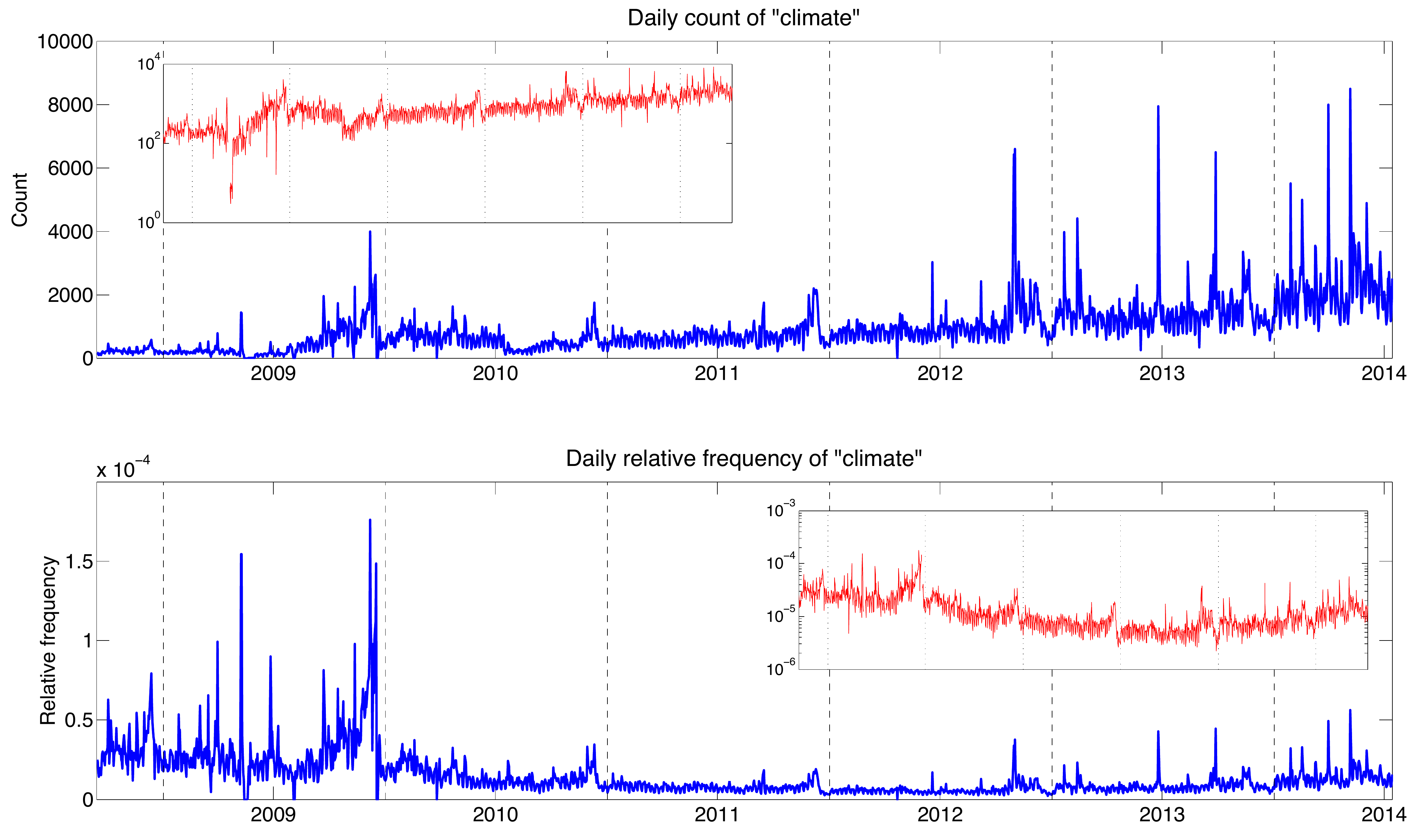}
\caption{The daily raw frequencies (top) and relative frequencies (bottom) of the word "climate" on Twitter from September 14, 2008 to July 14, 2014.  The insets (in red) show the same quantity with a logarithmically spaced y-axis.}
\label{freq}
\end{figure*}
\begin{figure*}[tbp!]
\includegraphics[width=0.9\textwidth]{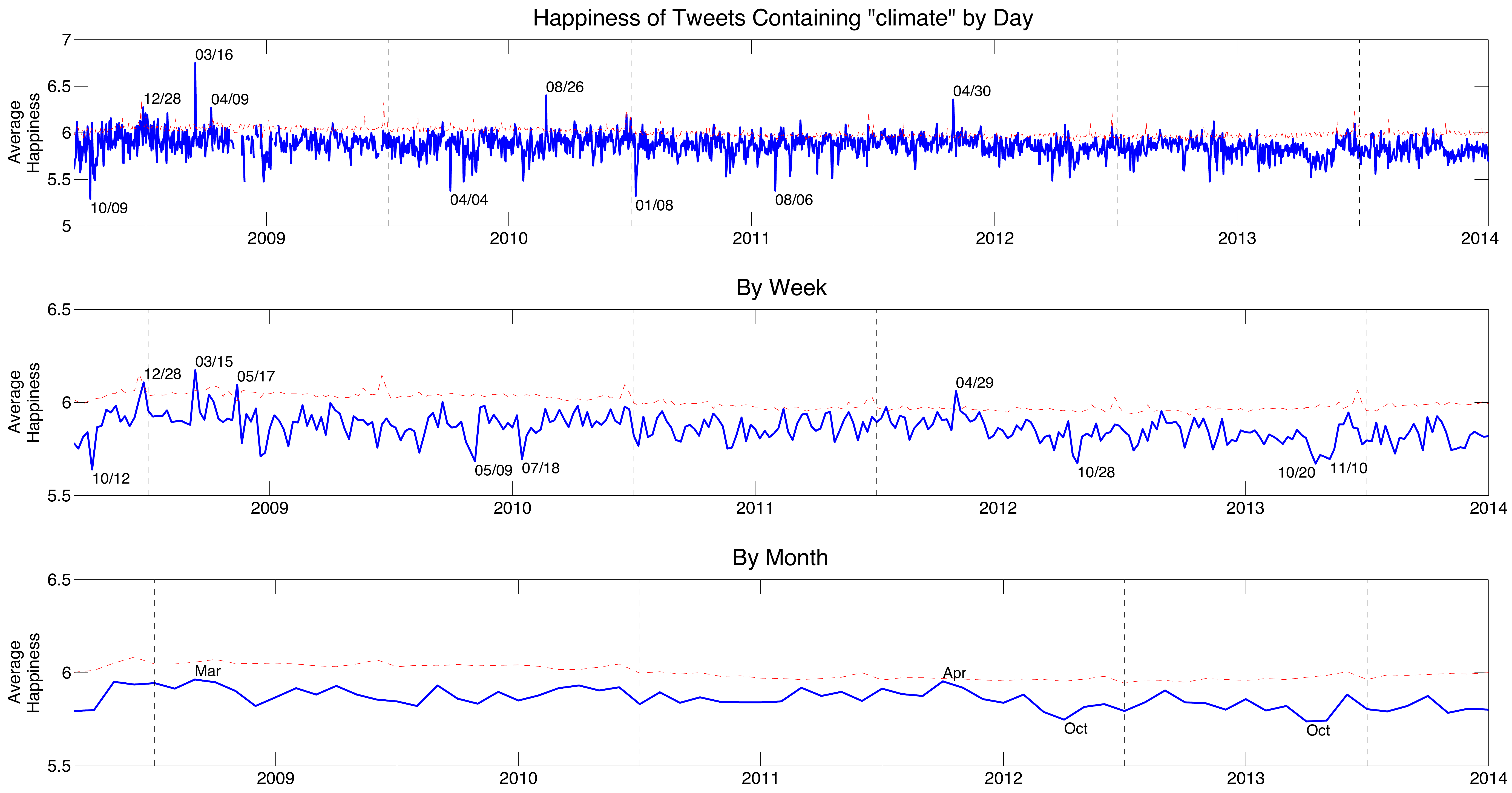}
\caption{Average happiness of tweets containing the word "climate" from September 2008 to July 2014 by day (top), by week (middle), and by month (bottom).  The average happiness of all tweets during the same time period is shown with a dotted red line.  Several of the happiest and saddest dates are indicated on each plot, and are explored in subsequent figures.}
\label{happ}
\end{figure*}
\begin{figure*}[btp!]
\includegraphics[width=0.9\textwidth]{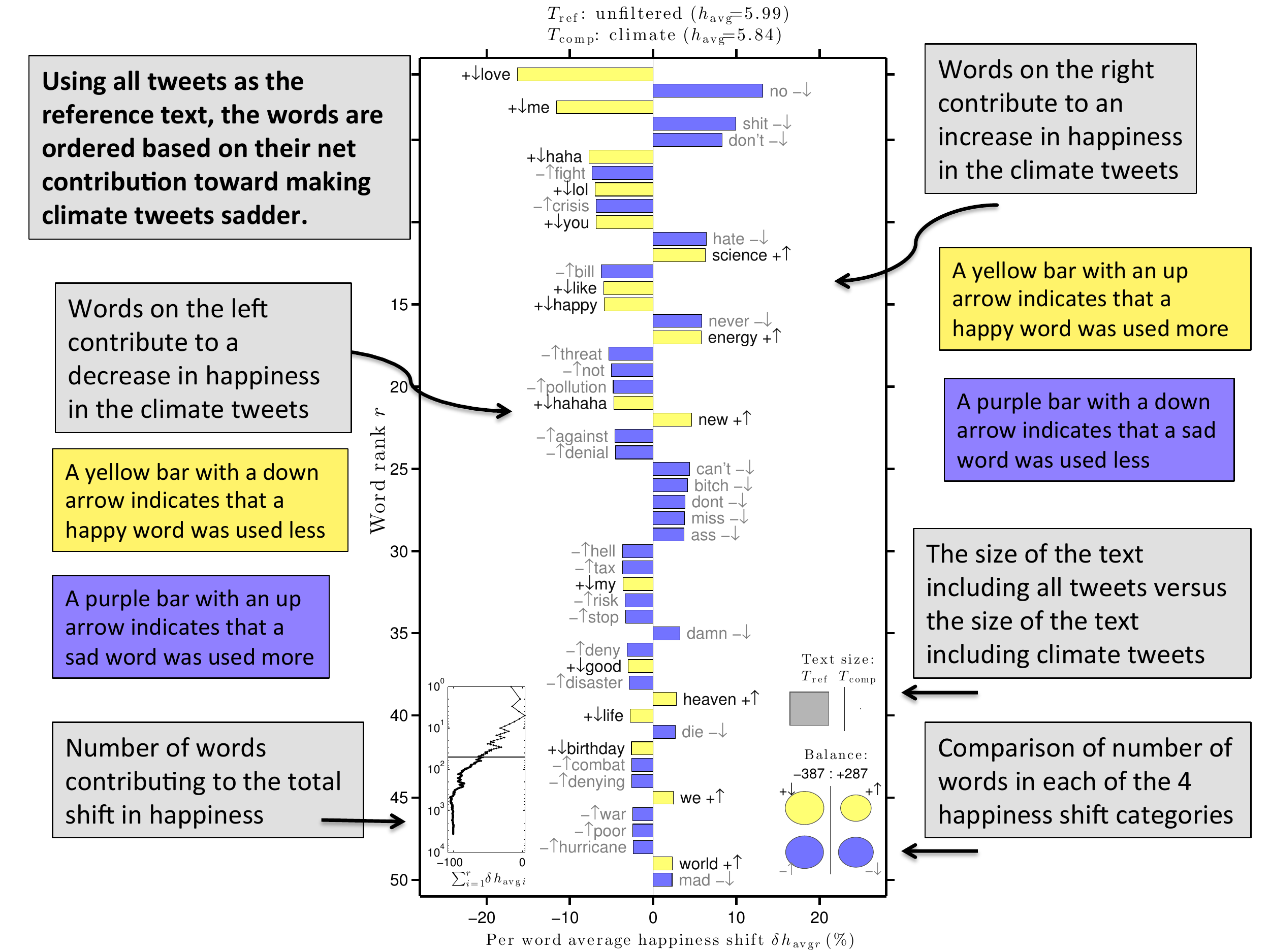}
\caption{A word shift graph comparing the happiness of tweets containing the word "climate" to all unfiltered tweets.  The reference text is roughly 100 billion tweets from September 2008 to July 2014.  The comparison text is tweets containing the word "climate" from September 2008 to July 2014.  A yellow bar indicates a word with an above average happiness score.  A purple bar indicates a word with below average happiness score.  A down arrow indicates that this word is used less within tweets containing the word "climate".  An up arrow indicates that this word is used more within tweets containing the word "climate".  Words on the left side of the graph are contributing to making the comparison text (climate tweets) less happy.  Words on the right side of the graph are contributing to making the comparison text more happy.  The small plot in the lower left corner shows how the individual words contribute to the total shift in happiness.  The gray squares in the lower right corner compare the sizes of the two texts, roughly $10^7$ vs $10^{12}$ words.  The circles in the lower right corner indicate how many happy words were used more or less and how many sad words were used more or less in the comparison text.  }
\label{climWS}
\end{figure*}

Fig.~\ref{freq} gives the raw and relative frequencies of the word "climate" over the study period.  We calculate the relative frequencies by dividing the daily count of "climate" by the daily sum of the 50,000 most frequently used words in the gardenhose sample.  From this figure, we can see that while the raw count increases over time, the relative frequency decreases over time.  This decrease can either be attributed to reduced engagement on the issue since the maximum relative frequency in December 2009, during Copenhagen Climate Change Conference, or an increase in overall topic diversity of tweets as Twitter grows in popularity.  The observed increase in raw count can largely be attributed to the growth of Twitter during the study period from approximately 1 million tweets per day in 2008 to approximately 500 million in 2014.  In addition, demographic changes in the user population clearly led to a decrease in the relative usage of the word "climate".

Fig.~\ref{happ} shows the average happiness of the climate tweets by day, by week, and by month during the 6 year time span.  The average happiness of unfiltered tweets is shown by a dotted red line.  Several high and low dates are indicated in the figure.  The average happiness of tweets containing the word "climate" is consistently lower than the happiness of the entire set of tweets. 

Several outlier days, indicated on the figure, do have an average happiness higher than the unfiltered tweets.  Upon recovering the actual tweets, we discover that on March 16, 2009, for example, the word "progress" was used 408 times in 479 overall climate tweets.  "Progress" has a happiness score of 7.26, which increases the average happiness for that particular day.  Increasing the time period for which the average happiness is measured (moving down the panels in Fig.~\ref{happ}), the outlier days become less significant, and there are fewer time periods when the climate tweets are happier than the reference tweets.  After averaging weekly and monthly happiness scores, we see other significant dates appearing as peaks or troughs in Fig.~\ref{happ}.  For example, the week of October 28, 2012 appears as one of the saddest weeks for climate discussion on Twitter.  This is the week when Hurricane Sandy made landfall on the east coast of the U.S.  For the same reason, October 2012 also appears as one of the saddest months for climate discussion.  

The word shift graph in Fig.~\ref{climWS} shows which words contributed most to the shift in happiness between climate tweets and unfiltered tweets.  The total average happiness of the reference text (unfiltered tweets) is 5.99 while the total average happiness of the comparison text (climate tweets) is 5.84.  This change in happiness is due to the fact that many positively rated words are used less and many negatively rated words are used more when discussing the climate.

The word "love" contributes most to the change in happiness.  Climate change is not typically a positive subject of discussion, and tweets do not typically profess love for it.  Rather, people discuss how climate change is a "fight", "crisis", or a "threat".  All of these words contribute to the drop in happiness.  Words such as "pollution", "denial", "tax", and "war" are all negative, and are used relatively more frequently in climate tweets, contributing to the drop in happiness.  The words "disaster" and "hurricane" are used more frequently in climate tweets, suggesting that the subject of climate change co-occurs with mention of natural disasters, and strong evidence exists proving Twitter is a valid indicator of real time attention to natural disasters \cite{ripberger2014social}.

On the positive side, we see that relatively less profanity is used when discussing the climate, with the exception of the word "hell".  We also see that "heaven" is used more often.  From our inspection of the tweets, it is likely that these two words appear because of a famous quote by Mark Twain: "Go to heaven for the climate and hell for the company" \cite{twain1980mark}.  Of the 97 non-earth related climate tweets from our 1,500 tweet sample, 8 of them referenced this quote.  The word "energy" is also used more during climate discussions.  This indicates that there may be a connection between energy related topics and climate related topics.  As energy consumption and types of energy sources can contribute to climate change, it is not surprising to see the two topics discussed together.

Using the first half of our dataset, Dodds et. al. \cite{dodds2011temporal} calculated the average happiness of tweets containing several individual keywords including "climate". They found that tweets containing the word "climate" were, on average, similar in ambient happiness to those containing the words "no", "rain", "oil", and "cold" (see Table 2 \cite{dodds2011temporal}).  \tb{In the following section, we compare the happiness score of tweets containing the word "climate" to that of 5 other climate-related keywords.}

\subsection*{Climate Related Keywords}
\tb{The diction used to describe climate change attitudes on Twitter may vary by user.  For example, some users may consistently use "climate change" and others may use "global warming".  There are also cohorts of users that utilize various hashtags to express their climate change opinions.  In order to address this, we collected tweets containing 5 other climate related keywords to explore the variation in sentiment surrounding different types of climate related conversation.  As in \cite{williams2015network}, we choose to analyze the keywords "global warming" (5.72), "globalwarming" (5.81), "climaterealists" (5.79), "climatechange" (5.86), and "agw" (5.73, standing for "anthropogenic global warming").  Search terms lack spaces in the cases where they are climate related hashtags.}  

\tb{Tweets including the "global warming" keyword contain more negatively rated words than tweets including "climate".  There is more profanity within these tweets and there are also more words suggesting that climate change deniers use the term "global warming" more often than "climate change".  For example, there is more usage of the words "stop", "blame", "freezing", "fraud", and "politicians" in tweets containing "global warming".  These tweets also show less frequent usage of positive words "science" and "energy", indicating that climate change science is discussed more within tweets containing "climate".  We also see a decrease in words such as "crisis", "bill", "risk", "denial", "denying", "disaster",  and  "threat".  The positively rated words "real" and "believe" appear more in "global warming" tweets, however so does the word "don't", again indicating that in general, the Twitter users who who don?t acknowledge climate change use the term "global warming" more frequently than "climate change".  A study in 2011 determined that public belief in climate change can depend on whether the question uses "climate change" or "global warming" \cite{schuldt2011global}.}  

\tb{Tweets containing the hashtag "globalwarming" also contain words indicating that this is often a hashtag used by deniers.  The word contributing most to the decrease in happiness between "climate" and "globalwarming" is "fail", possibly referencing an inaccurate interpretation of the timescale of global warming consequences during cold weather.  We see an increase in negative words "fraud", "die", "lie", "blame", "lies", and again a decrease in positive, scientific words.  There is also an increase in several cold weather words including "snow", "freezing", "christmas", "december", indicating that the "globalwarming" hashtag may often be used sarcastically.  Similarly, Tweets including the hashtag "climaterealists" use more words like "fraud", "lies", "wrong", and "scandal" and less "fight", "crisis", "pollution", "combat", and "threat".}  

\tb{The hashtag "agw" represents a group that is even more so against anthropogenic climate change.  We see an increase in "fraud", "lie", "fail", "wrong", "scare", "scandal", "conspiracy", "crime", "false", and "truth".  This particular hashtag gives an increase in positive words "green" and "science", however based on the large increase in the aforementioned negative words, we can deduce that these terms are being discussed in a negative light.  The "climatechange" hashtag represents users who are believers in climate change.  There is an increase in positive words "green", "energy", "environment", "sea", "oceans", "nature", "earth", and "future", indicating a discussion about the environmental impacts of climate  change.  There is also an increase in "pollution", "threat", "risk", "hunger", "fight", and "problem" indicating that the "climatechange" hashtag is often used when tweeting about the fight against climate change.}  

\tb{With the exception of the "globalwarming" hashtag, our analysis of these keywords largely agrees with what is found in \cite{williams2015network}.  Our analysis, however, compares word frequencies within tweets containing these hashtags with word frequencies within tweets containing the word "climate".  We find that more skeptics use "global warming" in their tweets than "climate", while it may be the case that "global warming" and "globalwarming" hashtag are also used by activists.}
\begin{figure*}[tph!]
\includegraphics[width=0.93\textwidth]{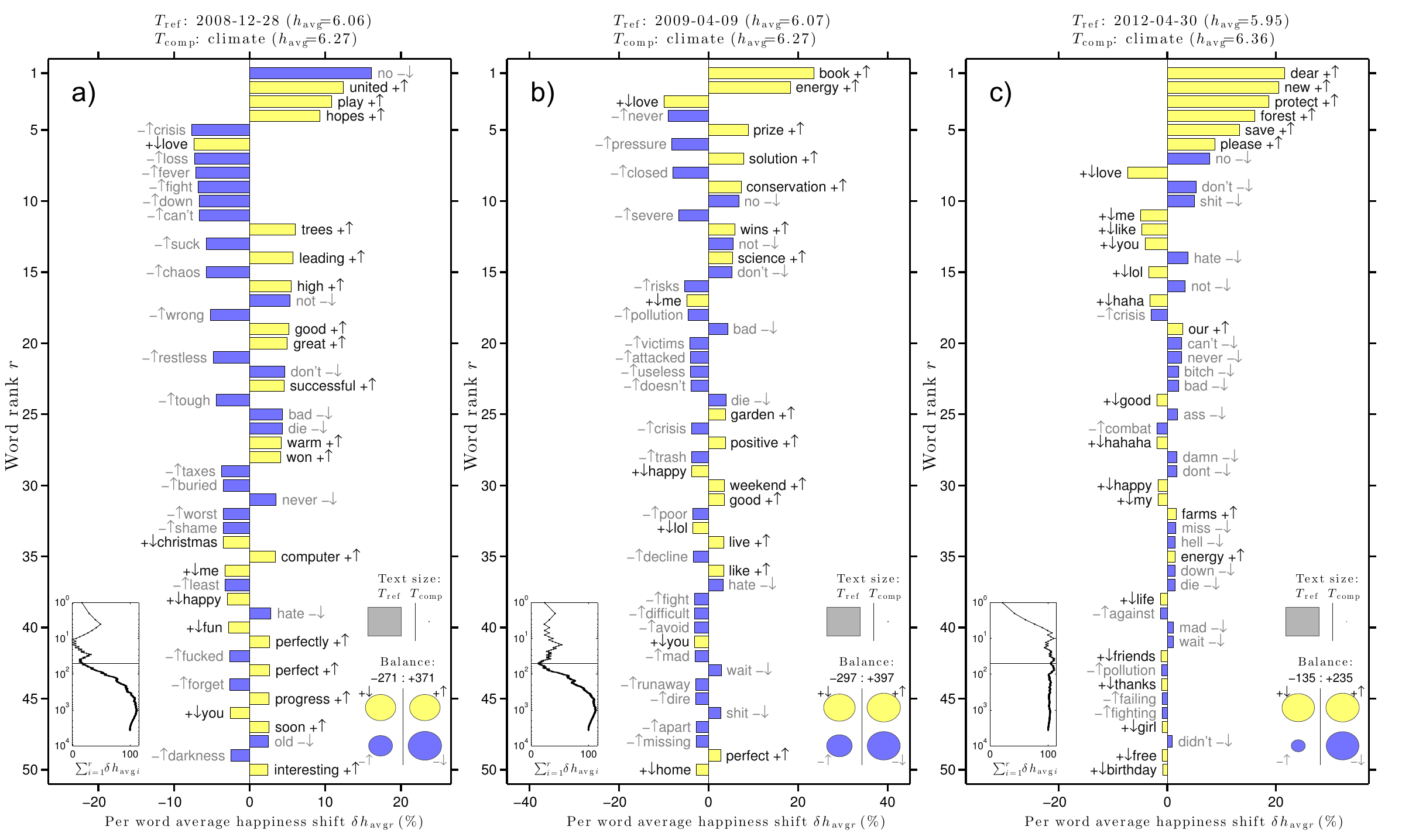}
\caption{Word shift graphs for three of the happiest days in the climate tweet time series.}
\label{daysH}
\end{figure*}
\subsection*{Analysis of Specific Dates}
\begin{figure}[tp!]
\includegraphics[width=0.42\textwidth]{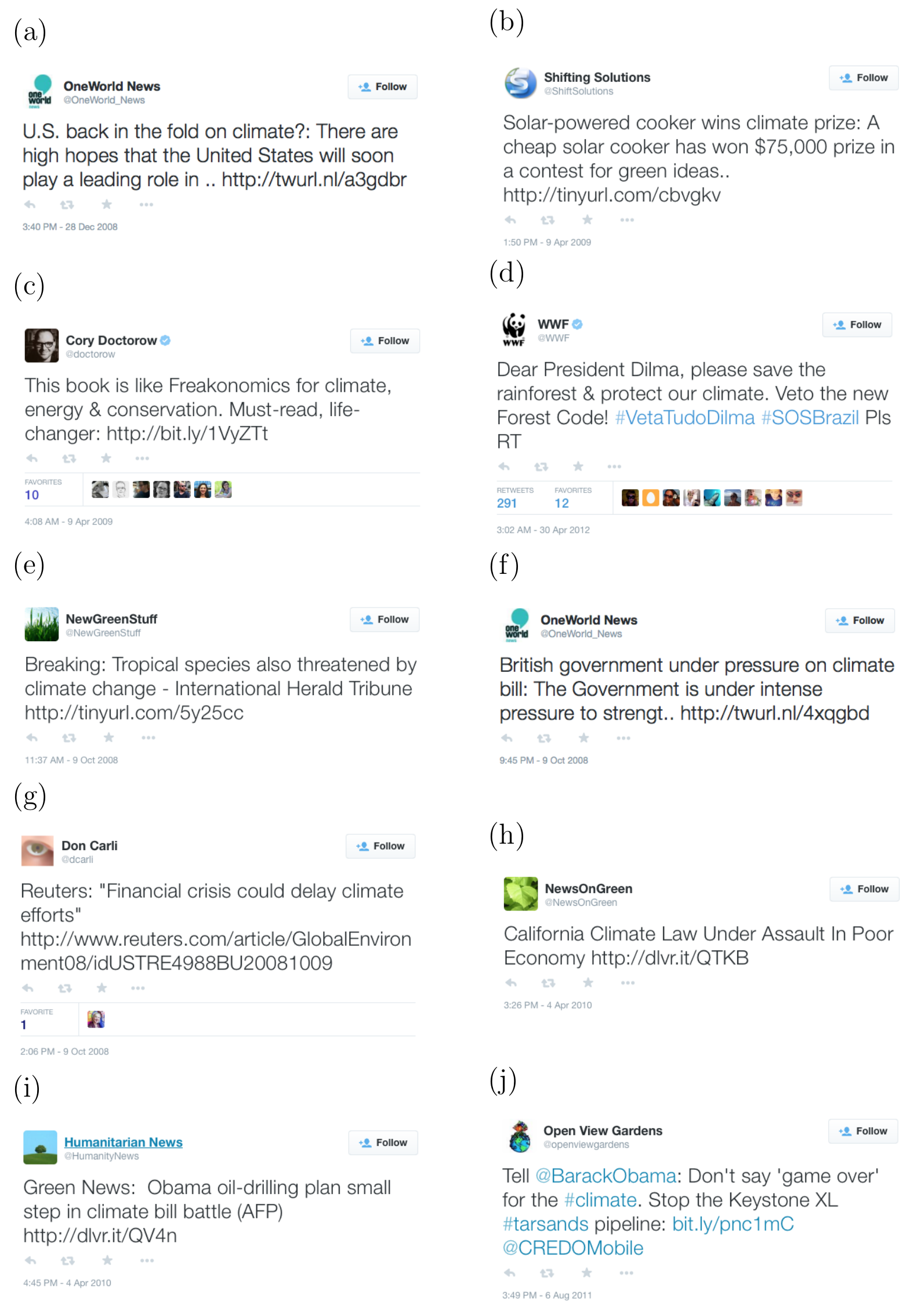}
\caption{Example tweets on the happiest and saddest days for climate conversation on Twitter}
\label{tweets}
\end{figure}
While Fig.~\ref{climWS} shows a shift in happiness for all climate tweets collected in the 6 year period, we now move to analyzing specific climate change-related time periods and events that correspond to spikes or dips in happiness. \tb{It is important to note that tweets including the word "climate" represent a very small fraction of unfiltered tweets (see gray squares comparing text sizes in bottom right of Fig.~\ref{climWS}).  While our analysis may capture specific events pertaining to climate change, it may not capture everything, as Twitter may contain background noise that we can't easily analyze.}\tb

Fig.~\ref{daysH} gives word shift graphs for three of the happiest days according to the hedonometer analysis.  These dates are indicated in the top plot in Fig.~\ref{happ}. The word shift graphs use unfiltered tweets as the reference text and climate tweets as the comparison text for the date given in each title.  Fig.~\ref{daysH}(a) shows that climate tweets were happier than unfiltered tweets on December 28, 2008.  This is due in part to a decrease in the word "no", and an increase in the words "united", "play", and "hopes".  On this day, there were "high hopes" for  the U.S. response to climate change.  An example tweet by OneWorld News is given in Fig.~\ref{tweets}(a) \cite{oneworldnews}.
\begin{figure*}[tp!]
\includegraphics[width=\textwidth]{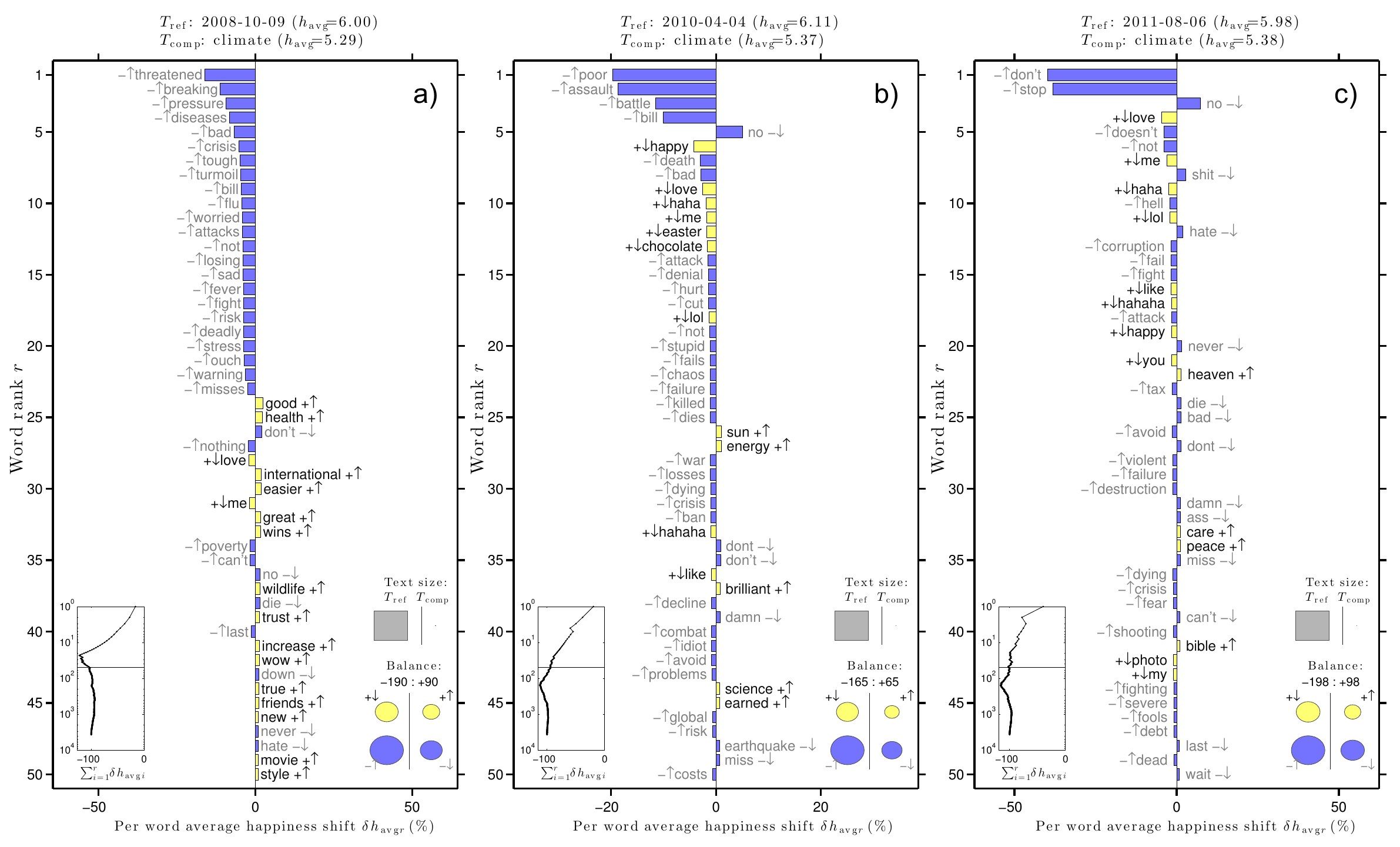}
\caption{Word shift graphs for 3 of the saddest days in the climate tweet time series.}
\label{daysS}
\end{figure*}

Fig.~\ref{daysH}(b) shows that climate tweets were happier than unfiltered tweets on April 9, 2009, largely due to the increase in positive words "book", "energy", and "prize".  Twitter users were discussing the release of a new book called \textit{Sustainable Energy Without the Hot Air} by David JC MacKay \cite{mackay2008sustainable}.  Also on this date, users were posting about a Climate Prize given to a solar-powered cooker in a contest for green ideas.  Example tweets include Fig.~\ref{tweets}(b) and (c)  \cite{solarcooker, book}.  Finally, Fig.~\ref{daysH}(c) shows that climate tweets were happier than unfiltered tweets on April 30, 2012.  This is due to the increased usage of the words "dear", "new", "protect", "forest", "save", and "please".  On this date, Twitter users were reaching out to Brazilian president Dilma to save the Amazon rainforest, e.g., Fig.~\ref{tweets}(d) \cite{wwf}.
 
Similarly, Fig.~\ref{daysS}  gives word shift graphs for three of the saddest days according to the hedonometer analysis.  These dates are indicated in the top panel in Fig.~\ref{happ}.  Fig.~\ref{daysS}(a) shows an increase in many negative words on October 9, 2008.  Topics of conversation in tweets containing "climate" include the threat posed by climate change to a tropical species, a British climate bill, and the U.S. economic crisis.  Example tweets include Fig.~\ref{tweets}(e-g) \cite{tropical, oneworldnews2, financial}.

Fig.~\ref{daysS}(b) shows an increase in negative words "poor","assault", "battle", and "bill" on April 4, 2010.  Popular topics of conversation on this date included a California climate law and President Obama's oil-drilling plan.  Example tweets include Fig.~\ref{tweets}(h) and (i) \cite{cali,drilling}.  Finally, Fig.~\ref{daysS}(c) shows that the words "don't" and "stop" contributed most to the decrease in happiness on August 6, 2011.  A topic of conversation on this date was the Keystone XL pipeline, a proposed extension to the current Keystone Pipeline.  An example tweet is given in Fig.~\ref{tweets}(j) \cite{obama}.  
 
This per day analysis of tweets containing "climate" shows that many of the important issues pertaining to climate change appear on Twitter, and demonstrate different levels of happiness based on the events that are unfolding.  In the following section, we investigate specific climate change events that may exhibit a peak or a dip in happiness.  First, we analyze the climate change discussion during several natural disasters that may have raised awareness of some of the consequences of climate change.  Then, we analyze a non-weather related event pertaining to climate change.

\subsection*{Natural Disasters}

\begin{figure}[tbp!]
\includegraphics[width=0.5\textwidth]{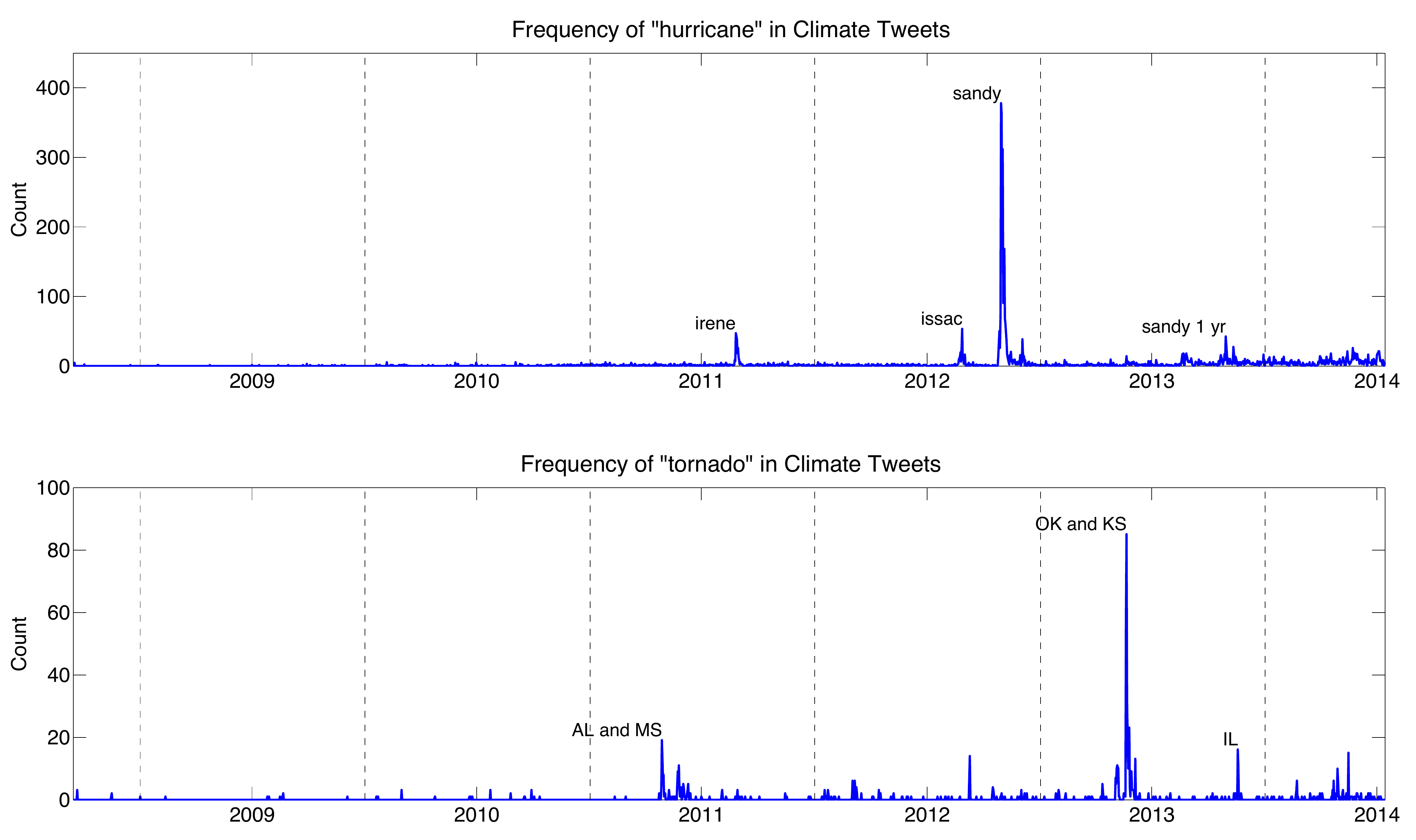}
\caption{Frequency of the word "hurricane'" (top) and "tornado" (bottom) within tweets containing the word "climate".  Several spikes have been identified with the hurricane or tornado that took place during that time period.}
\label{hurricane}
\end{figure}

Natural disasters such as hurricanes and tornados have the potential to focus society's collective attention and spark conversations about climate change.  A person's belief in climate change is often correlated with the weather on the day the question is asked \cite{zaval2014warm,li2011local,hamilton2013blowin}.  A study using "climate change" and "global warming" tweets showed that both weather and mass media coverage heavily influence belief in anthropogenic climate change \cite{kirilenko2015people}.  In this section, we analyze tweets during three natural disasters: Hurricane Irene, Hurricane Sandy, and a midwest tornado outbreak that damaged many towns including Moore, Oklahoma and Rozel, Kansas.  Fig.~\ref{hurricane} gives the frequencies of the words "hurricane" and "tornado" within tweets that contain the word "climate".  Each plot labels several of the spikes with the names of the hurricanes (top) or the locations (state abbreviations) of the tornado outbreaks (bottom).  This figure indicates that before Hurricane Irene in August 2011, hurricanes were not commonly referenced alongside climate, and before the April 2011 tornado outbreak in Alabama and Mississippi, tornados were not commonly referenced alongside climate.

\tb{This analysis, however, will not capture every hurricane or tornado mentioned on Twitter, only those that were referenced alongside the word "climate". Hurricane Arthur, for example, occurred in early July, 2014 and does not appear as a spike in Fig.~\ref{hurricane}. This particular hurricane did not cause nearly as much damage or as many fatalities as the hurricanes that do appear in Fig.~\ref{hurricane}, and perhaps did not draw enough attention to highlight a link between hurricanes and climate change on Twitter. Additionally, a large tornado outbreak in Kentucky, Alabama, Indiana, and Ohio occurred in early March 2012 and does not appear as a spike in our analysis.}\tb

\begin{figure}[tbp!]
\includegraphics[width=0.5\textwidth]{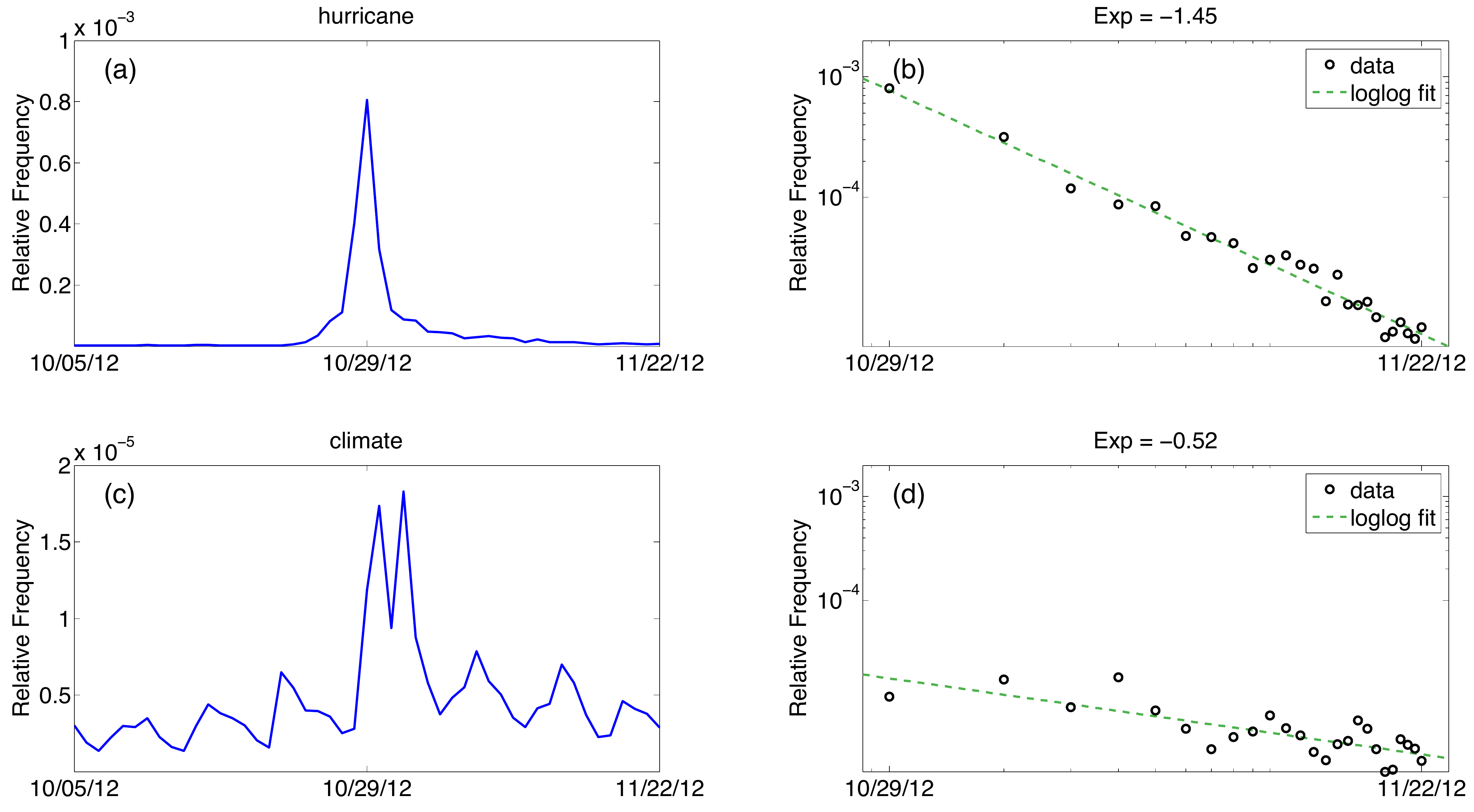}
\caption{Decay rates of the words "hurricane" (top) and "climate" (bottom).  The left plots gives the time series of each word during hurricane Sandy.  The right plots gives the power law fit for the decay in relative frequency, x-axes are spaced logarithmically.  The power law exponents are given in the titles of the figures.}
\label{decay}
\end{figure}

Fig.~\ref{hurricane} shows that the largest peak in the word "hurricane" occurred during Hurricane Sandy in October 2012.  Fig.~\ref{decay} provides a deeper analysis for the climate time series during hurricane Sandy.  The time series of the words "hurricane" and "climate" as a fraction of all tweets before, during, and after Hurricane Sandy hit are given in Fig.~\ref{decay}(a) and (c).  Spikes in the frequency of usage of these words is evident in these plots.  The decay of each word is fitted with a power law in Fig.~\ref{decay}(b) and (d).  A power law is a functional relationship of the following form:
\begin{equation}
f(t-t_{event}) = \alpha(t-t_{event})^{-\gamma}
\label{power}
\end{equation}
Here, $t$ is measured in days, and $t_{event}$ is the day Hurricane Sandy made landfall.  $f(t)$ represents the relative frequency of the word "hurricane" (top) or "climate" (bottom), and $\alpha$ and $\gamma$ are constants.

\begin{figure*}[tbp!]
\includegraphics[width=\textwidth]{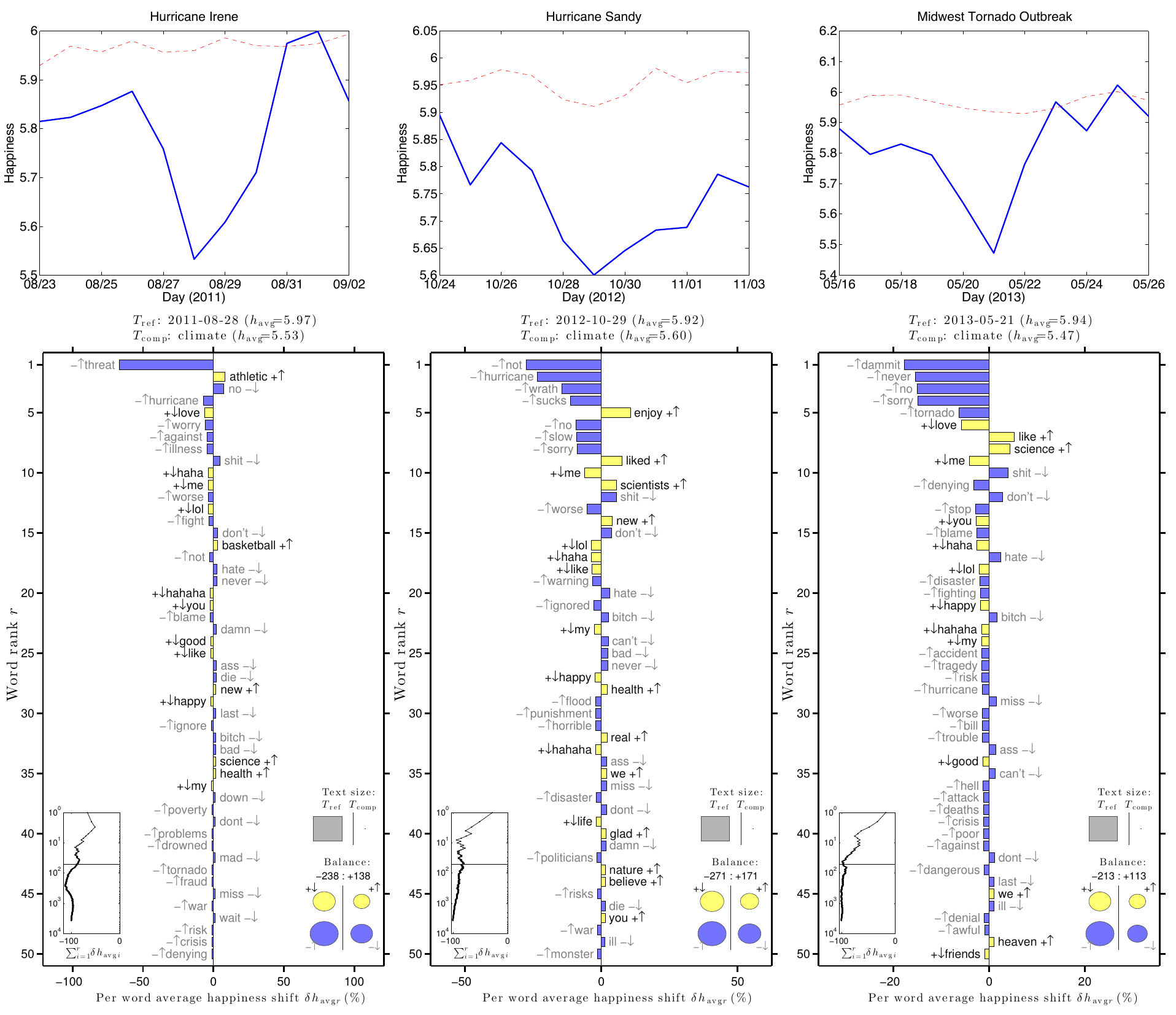}
\caption{Happiness time series plots for tweets containing the word "climate" one week before and one week after three natural disasters in the United States (top) and word shift graphs indicating what words contributed most to the drop in happiness during the natural disasters (bottom).  The word shift graphs compare the climate tweets to unfiltered tweets on the day of the natural disaster.}
\label{disasters}
\end{figure*}

Using the power law fit, we calculate the first three half lives of the decay.  Letting $M$ equal the maximum relative frequency, the time at which the first half life of the power law relationship occurs is calculated by equation \ref{half}:
\begin{equation}
t_{\frac{1}{2}} = \left(\frac{M}{2\alpha}\right)^{-\frac{1}{\gamma}}
\label{half}
\end{equation}
The first three half lives of the decay in the frequency of the word "hurricane" during hurricane Sandy are 1.57, 0.96, and 1.56 additional days.  Since the decay is not exponential, these half lives are not constant.  The first half life indicates that after about a day and a half, "hurricane" was already tweeted only half as often.  The second half life indicates that after one more day, "hurricane" was tweeted only one fourth as often, and so on.  Thus, it did not take long for the discussion of the hurricane to decrease.  The half lives, however, of the word "climate" are much larger at  8.19, 22.58, and 84.85 days.

Fig.~\ref{disasters} gives happiness time series plots for three natural disasters occurring in the United States.  These plots show that there is a dip in happiness on the day that the disasters hit the affected areas, offering additional evidence that sentiment is depressed by natural disasters \cite{antracking}.  The word shift graphs indicate which words contributed to the dip in happiness.  The circles on the bottom right of the word shift plots indicate that for all three disasters,  the dip in happiness is due to an increase in negative words, more so than a decrease in positive words.  During a natural disaster, tweets mentioning the word "climate" use more negative words than tweets not mentioning the word "climate".

\subsection*{Forward on Climate Rally}

\begin{figure}[h]
\includegraphics[width=0.5\textwidth]{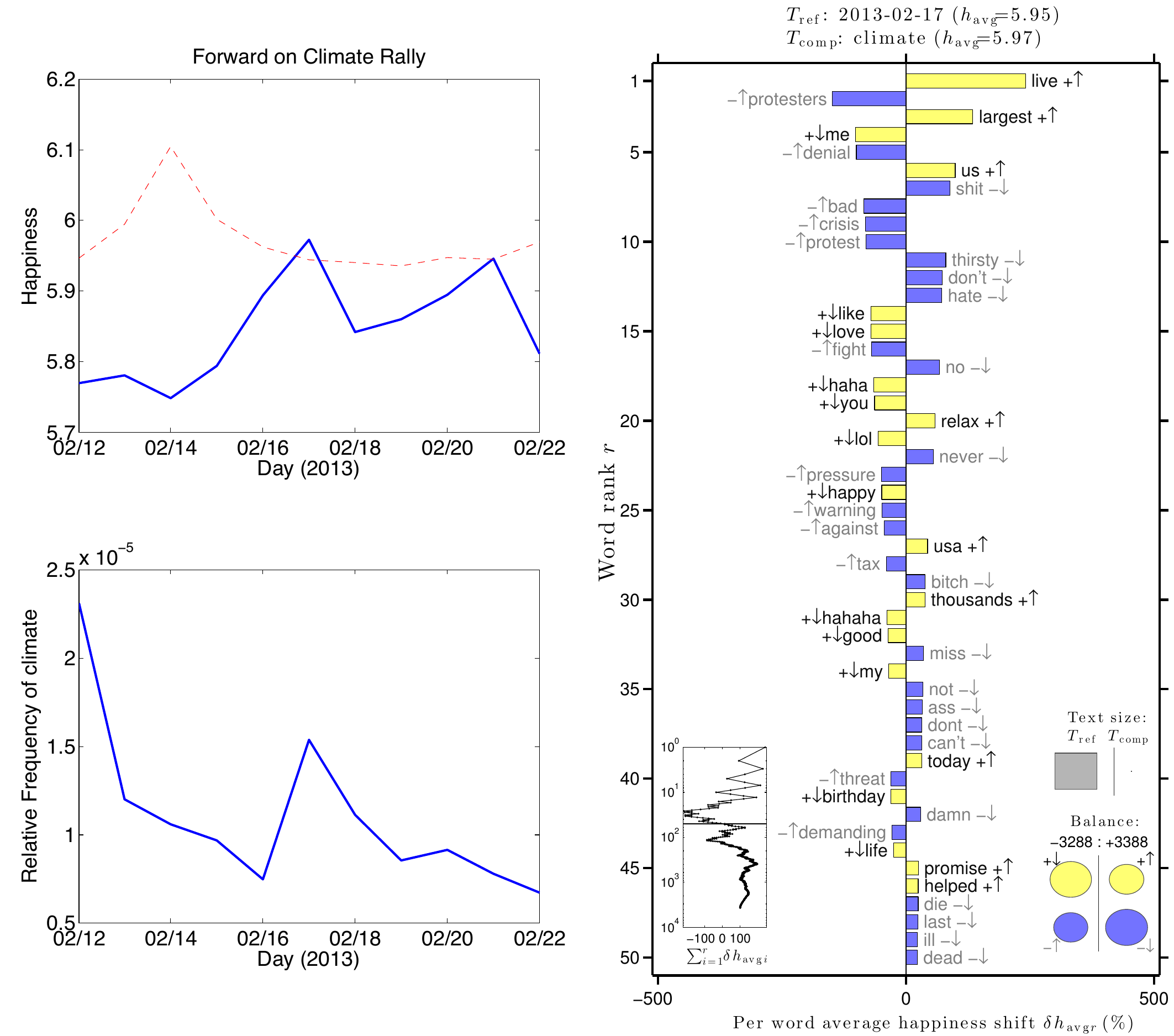}
\caption{Left: Happiness time series plot for unfiltered tweets (red dashed) and tweets containing the word "climate" (blue solid) one week before and one week after the Forward on Climate Rally.  Right: word shift plot for climate tweets versus unfiltered tweets on the day of the rally.}
\label{rally}
\end{figure}

In this section, we analyze tweets during the Forward on Climate Rally, which took place in Washington D.C. on February 17, 2013.  The goal of the rally, one of the largest climate rallies ever in the United States, was to convince the government to take action against climate change.  The proposed Keystone pipeline bill was a particular focus.  Fig.~\ref{rally} shows that the happiness of climate tweets increased slightly above the unfiltered tweets during this event, which only occurs on 8\% of days in Fig.~\ref{happ}.

Despite the presence of negative words such as "protestors", "denial", and "crisis", the Forward on Climate Rally introduced positive words such as "live", "largest", and "promise".  The Keystone pipeline bill was eventually vetoed by President Obama. 

\section*{Conclusion}
We have provided a general exploration of the sentiment surrounding tweets containing the word "climate" in response to natural disasters and climate change news and events.  The general public is becoming more likely to use social media as an information source, and discussion on Twitter is becoming more commonplace.  We find that tweets containing the word "climate" are less happy than all tweets.  In the United States, climate change is a topic that is heavily politicized; the words "deny", "denial", and "deniers" are used more often in tweets containing the word "climate".  The words that appear in our climate-related tweets word shift suggest that the discussion surrounding climate change is dominated by climate change activists rather than climate change deniers, indicating that the twittersphere largely agrees with the scientific consensus on this issue.  The presence of the words "science" and "scientists" in almost every word shift in this analysis also strengthens this finding (see also \cite{antracking}).  The decreased "denial" of climate change is evidence for how a democratization of knowledge transfer through mass media can circumvent the influence of large stakeholders on public opinion. 

In examining tweets on specific dates, we have determined that climate change news is abundant on Twitter.  Events such as the release of a book, the winner of a green ideas contest, or a plea to a political figure can produce an increase in sentiment for tweets discussing climate change.  For example, the Forward on Climate Rally demonstrates a day when the happiness of climate conversation peaked above the background conversation.  On the other hand, consequences of climate change such as threats to certain species, extreme weather events, and climate related legislative bills can cause a decrease in overall happiness of the climate conversation on Twitter due to an increase in the use words such as "threat", "crisis", and "battle".

Natural disasters are more commonly discussed within climate-related tweets than unfiltered tweets, implying that some Twitter users associate climate change with the increase in severity and frequency of certain natural disasters \cite{mann2006atlantic, huber2011extreme, field2012managing}.  During Hurricane Irene, for example, the word "threat" was used much more often within climate tweets, suggesting that climate change may be perceived as a bigger threat than the hurricane itself.  The analysis of Hurricane Sandy in Fig.~\ref{decay} demonstrates that while climate conversation peaked during Hurricane Sandy, it persisted longer than the conversation about the hurricane itself.

While climate change news is prevalent in traditional media, our research provides an overall analysis of climate change discussion on the social media site, Twitter.  Through social media, the general public can learn about current events and display their own opinions about global issues such as climate change.  Twitter may be a useful asset in the ongoing battle against anthropogenic climate change, as well as a useful research source for social scientists, an unsolicited public opinion tool for policy makers, and public engagement channel for scientists.   

\acknowledgments
The authors are grateful for the computational resources provided by the Vermont Advanced Computing Core which is supported by the Vermont Complex Systems Center.  CMD, AJR, EMC, and LM were supported by National Science Foundation (NSF) grant DMS-0940271 to the Mathematics \& Climate Research Network.  PSD was supported by NSF CAREER Award \#0846668.  Thanks to Mary Lou Zeeman for helpful discussions.

\bibliographystyle{unsrt}


\clearpage

\end{document}